\documentclass[twocolumn,pre,nofootinbib,floatfix,superscriptaddress]{revtex4}

\usepackage{dcolumn}
\usepackage{graphicx}
\usepackage{rotating}
\usepackage{amsmath,amsfonts,amssymb}

\newcommand{\etal}{\emph{et al.\ }}
\newcommand{\etalp}{\emph{et al}}

\newcommand{\remove}[1]{}

\begin{document}

\title{Comment on Yu et al., ``High Quality Binary Protein Interaction Map of the Yeast Interactome Network.'' {\em Science} {\bf 322}, 104 (2008).}
\author{A. Clauset}
\email{aaronc@santafe.edu}
\affiliation{Santa Fe Institute, 1399 Hyde Park Road, Santa Fe NM, 87501, USA}

\begin{abstract}
We test the claim by Yu \etal---presented in {\em Science} {\bf 322}, 104 (2008)---that the degree distribution of the yeast ({\em Saccharomyces cerevisiae}) protein-interaction network is best approximated by a power law. Yu \etal consider three versions of this network. In all three cases, however, we find the most likely power-law model of the data is distinct from and incompatible with the one given by Yu \etalp. Only one network admits good statistical support for any power law, and in that case, the power law explains only the distribution of the upper 10\% of node degrees. These results imply that there is considerably more structure present in the yeast interactome than suggested by Yu \etalp., and that these networks should probably not be called ``scale free.''
\end{abstract}

\maketitle

Protein-interaction networks, where nodes are natural proteins and links represent non-trivial binding affinity between two proteins, hold great promise for pushing forward our understanding of cellular processes. The wide interest in these interactome networks stems mainly from the observation that while modern genetic methods allow us to identify which genes code for proteins, the set of these genes only amounts to a ``parts list'' of a cell. A more complete understanding of cellular processes requires knowing something about the functional roles and dynamic interactions of these parts~\cite{footnote:1}. Thus, by considering the patterns of protein interactions, i.e., their network, we can pose and answer meaningful biological questions about complex cellular processes, and their evolution. 

Determining the protein-interaction network for a particular species is a highly non-trivial task, and relies upon sophisticated molecular techniques to both build the proteins, and to test their pairwise interactions. The most direct approach would be to individually test each of the $n^{2}$ possible interactions for $n$ proteins. But, because $n$ is typically on the order of thousands or tens of thousands, and high-throughput methods are not yet available for these tests, this approach is not used. Instead, researchers use techniques that test multiple interactions at once~\cite{aebersold:mann:2003,phizicky:etal:2003}, e.g., the yeast two-hybrid (Y2H) screen~\cite{mcalisterhenn:etal:1999}. To date, there have been a number of high profile efforts to construct the interactome for yeast \mbox{({\em S. cerevisiae})}~\cite{fromont-racine:etal:1997,uetz:etal:2000,ito:etal:2001,gavin:etal:2002,ho:etal:2002}, and, many would argue, a lot of real progress.



However, these methods also have serious limitations, and have been shown capable of producing high false-positive and high false-negative rates~\cite{bader:hogue:2002,bader:etal:2004}. Some of the techniques also exhibit severe biases, being unable to test for interactions involving entire classes of proteins. The Y2H assay, for instance, is not suitable for transcriptional activators or membrane-bound proteins. As a result of these limitations, some scientists have wondered privately how much real biology our current maps actually capture.

The Yu \etal paper attempts to address some of these shortcomings using a new high-throughput Y2H screen. Combining its results with those of previous studies, they produce a new set of interactions (denoted Y2H-union), 
which they suggest covers about 20\% of the entire network. In the interest of completeness, Yu \etal include in their subsequent network analyses two alternative versions of the yeast interactome: one is based on co-complex information drawn from raw high-throughput coaffinity purification and mass spectrometry data (denoted Combined-AP/MS), and one is based on a smaller set of literature-curated interactions (denoted LC-multiple) that are assumed to be error free.

This Comment is not concerned with the quality of the laboratory procedures or the accuracy of the inferred interaction data. Rather, the focus is relatively narrow, concerning only the analysis of these networks' degree distributions and the conclusions drawn thereby. In particular, Yu \etal claim that the degree distributions of all three networks follow power-law distributions (with parameters given in Table~\ref{table:1}):
\begin{quotation}
\noindent As found previously for other macromolecular networks, the connectivity or ``degree'' distribution of all three data sets is best approximated by a power-law.\enspace\cite{yu:etal:2008}
\end{quotation}
The implication thus being that the yeast proteome can be considered ``scale free'', with all that goes along with that label~\cite{li:etal:2006}. 
However, this claim depends on a statistical method---specifically, linear regression---that is known to produce biased and incorrect results in this context. By reanalyzing Yu \etalp.'s data~\cite{footnote:2} with appropriate statistical tools~\cite{clauset:etal:2007}, we show that (i) the most likely power-law models of these networks' degree distribution are distinct from and incompatible with those quoted by Yu \etalp., (ii) at best, only the 10.3\% most connected nodes in the Y2H-union network are plausibly power-law distributed, and (iii) there is considerably more structure present in all of these networks than suggested by Yu \etal 
As a result, these networks should probably not be considered ``scale free.''

\begin{figure*}[t]
\begin{center}
\begin{tabular}{ccc}
\includegraphics[scale=0.31]{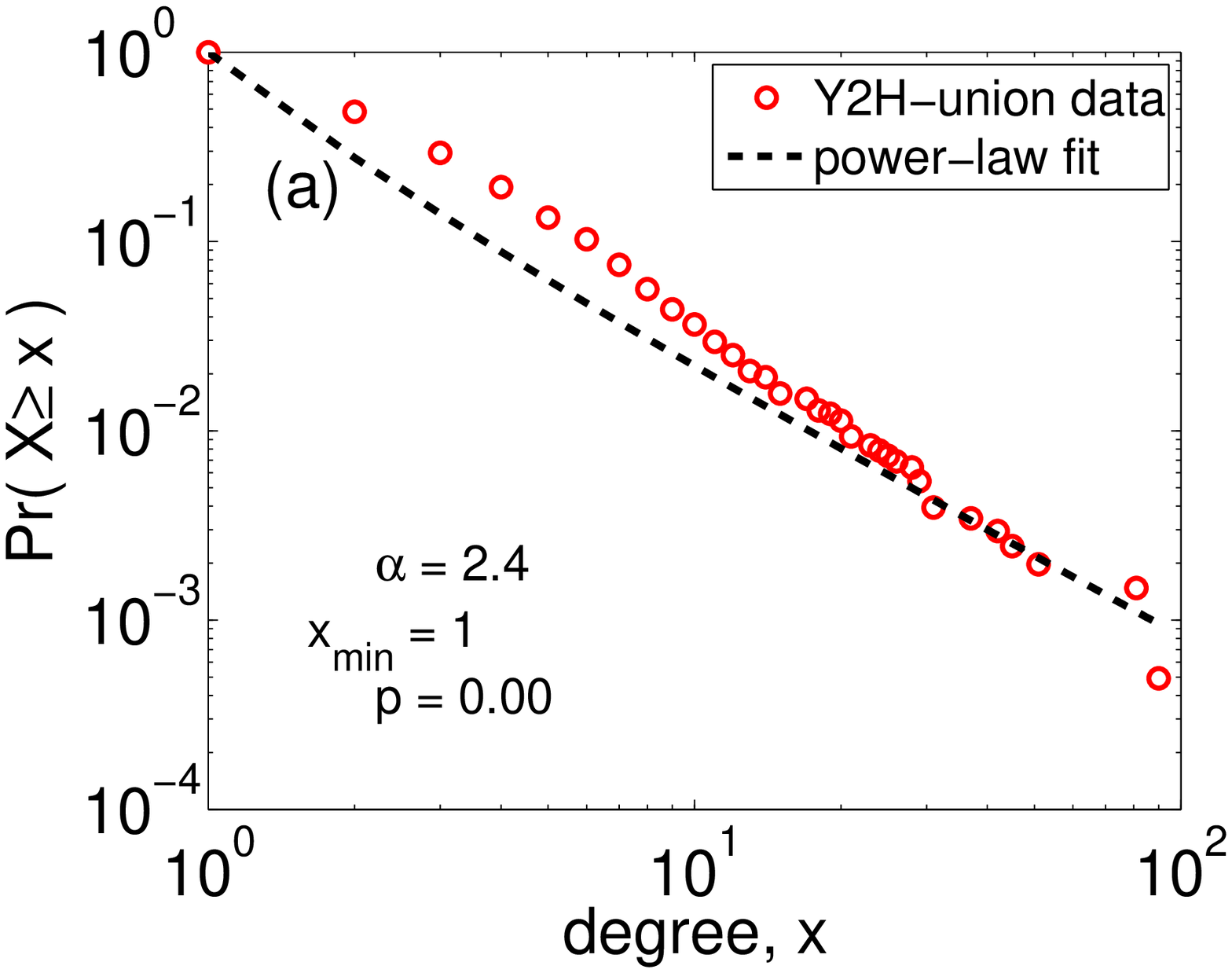} &
\includegraphics[scale=0.31]{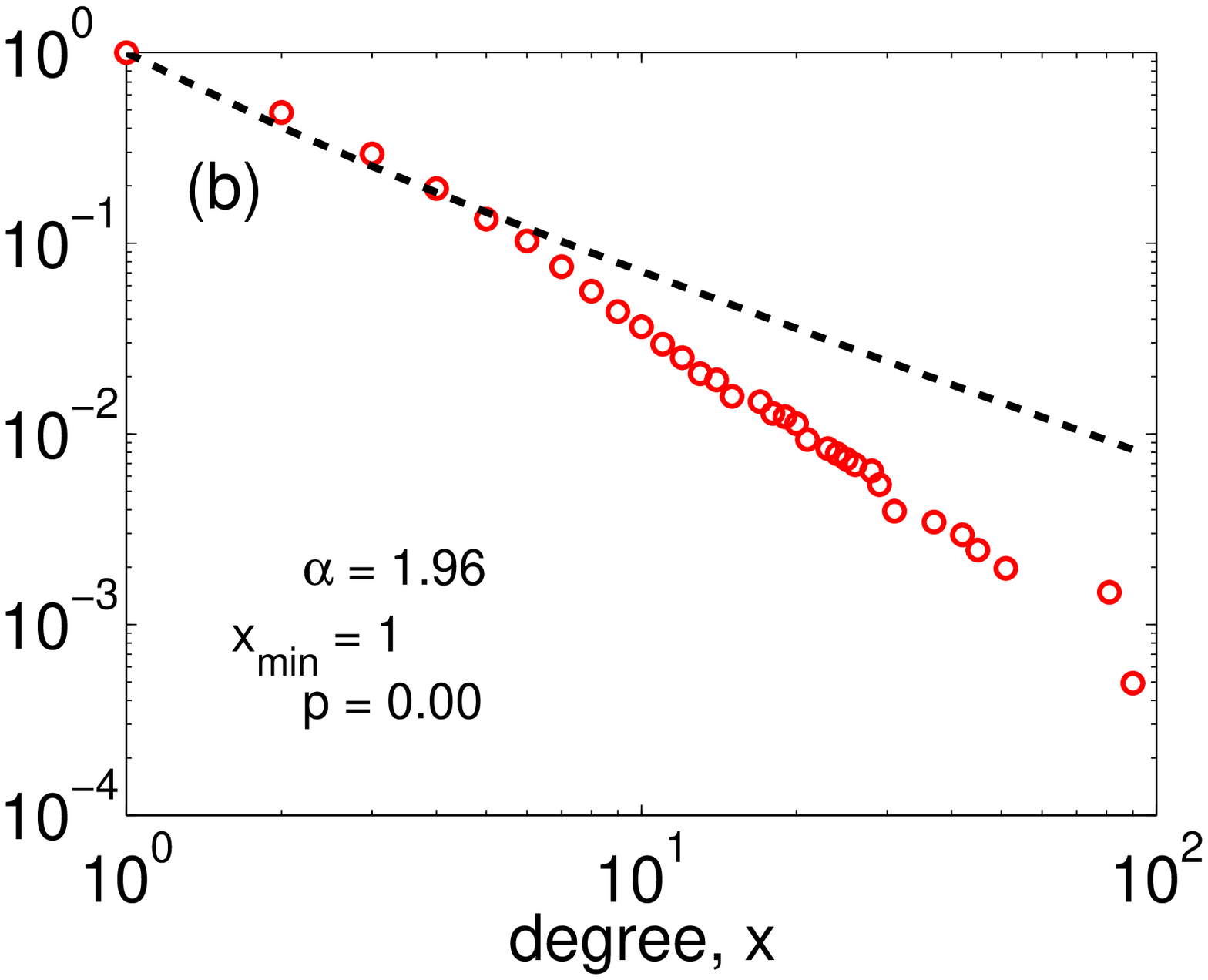} &
\includegraphics[scale=0.31]{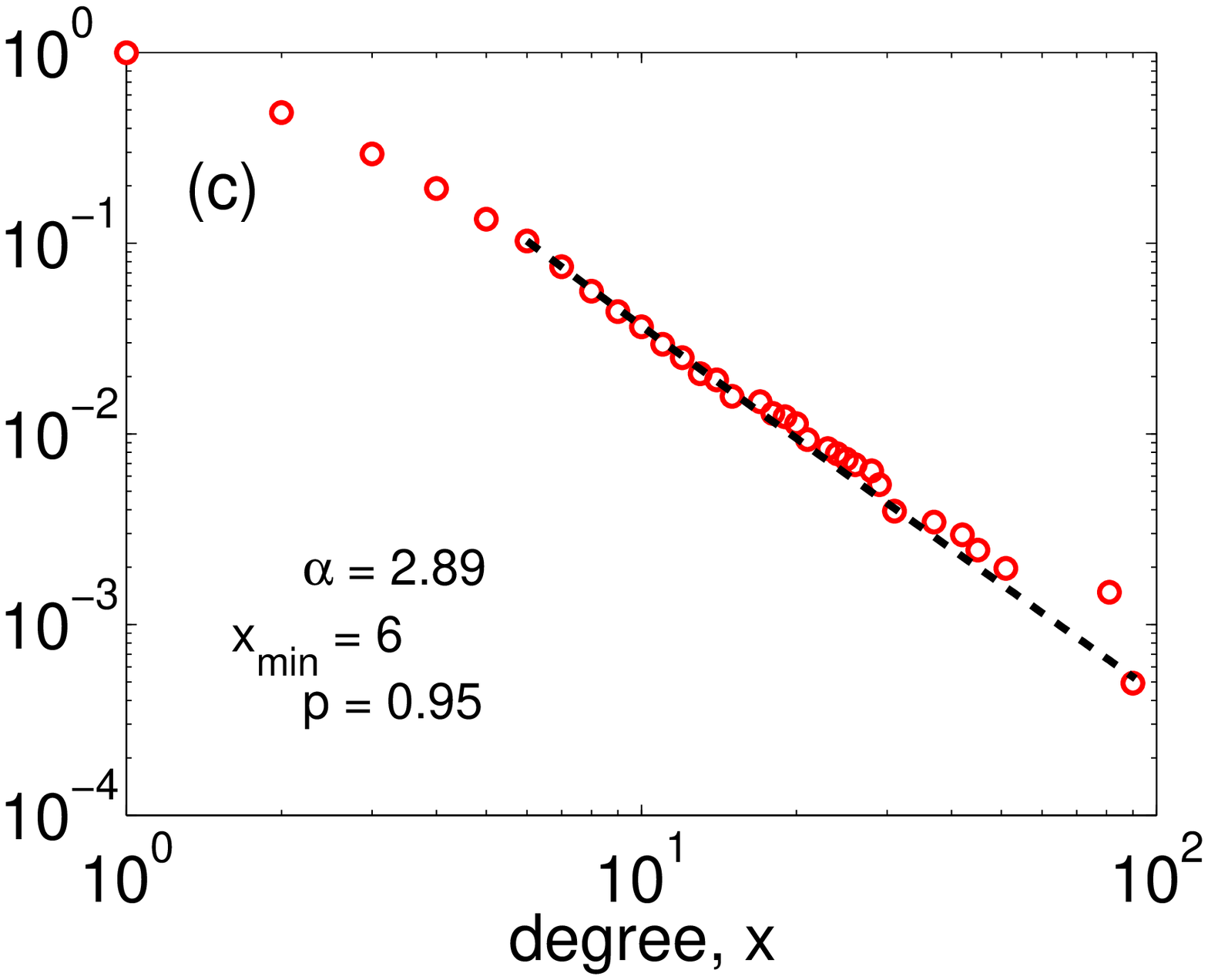} 
\end{tabular}
\end{center}
\caption{Three approaches to fitting a power-law distribution to the Y2H-union data from Yu \etal~\cite{yu:etal:2008}. (a) The parameterization given by Yu \etalp., derived using standard linear regression on the log-transformed histogram of degree frequencies. (b) A fit derived via maximum likelihood over the entire range of the data, i.e., $x_{\min} =1$. (c) A fit derived via maximum likelihood for estimating $\alpha$, and by selecting the $x_{\min}$ that gives the best power-law fit to the upper range of the data. In both (a) and (b), the fitted power-law is not statistically significant ($p=0.00\pm0.01$), indicating that large or systematic deviations from the power-law hypothesis exist. In (c), the upper 10.3\% are plausibly power-law distributed ($p=0.95\pm0.03$). }
\label{fig:S6A}
\end{figure*}

We begin by reanalyzing the Y2H-union data, argued by Yu \etal to be the most accurate map among the three. Figures~\ref{fig:S6A}a--c show the data along with three different power-law models. The first panel shows the model suggested by Yu \etal (with scaling exponent $\alpha=2.4$) which was derived using a linear regression approach; this model yields a poor fit to the lower-to-middle range of the data. The second panel shows the most likely power-law model over the entire range of data, which fits the lower range and thus the majority of the data relatively well, but yields a poor fit to the upper range. The third panel shows the most likely power-law model for the upper range alone.

Notably, Figures~\ref{fig:S6A}a--c plot the data as a complementary cumulative distribution function (CDF). If the data were indeed power-law distributed, this function would be straight on the log-log axes, but the reverse is not true. Being straight on log-log axes is not a sufficient condition for some data to be power-law distributed; many kinds of non-power-law distributed data can look straight on log-log axes. To decide whether some data do or do not follow a power-law distribution, we must use statistical tools that can tell the difference~\cite{clauset:etal:2007}. Linear regression is not one of these.

A straightforward test for whether some data can reasonably be claimed to follow a power-law distribution is a significance test~\cite{wasserman:2004,press:etal:1992}.
When we fit the power-law model to the same data that we use to score the power law's plausibility, we induce a correlation between the data and the model; however, this correlation can be controlled using a Monte Carlo procedure~\cite{clauset:etal:2007}. The result of the test is a single value $p$ that represents the plausibility of the fitted model as an explanation of the data: small $p$-values, conventionally $p<0.1$, indicate that the data cannot be considered to follow the fitted model, while larger values indicate only that the fitted model is plausible (not that it is correct)~\cite{footnote:5}. Conducting such a test with the Y2H-union data and the power-law claimed by Yu \etal yields $p=0.00\pm0.01$, indicating that this power law is a terrible model of the data, and that the data deviate in large or systematic ways from the proposed power law.

\begin{figure*}[t]
\begin{center}
\begin{tabular}{cc}
\includegraphics[scale=0.4]{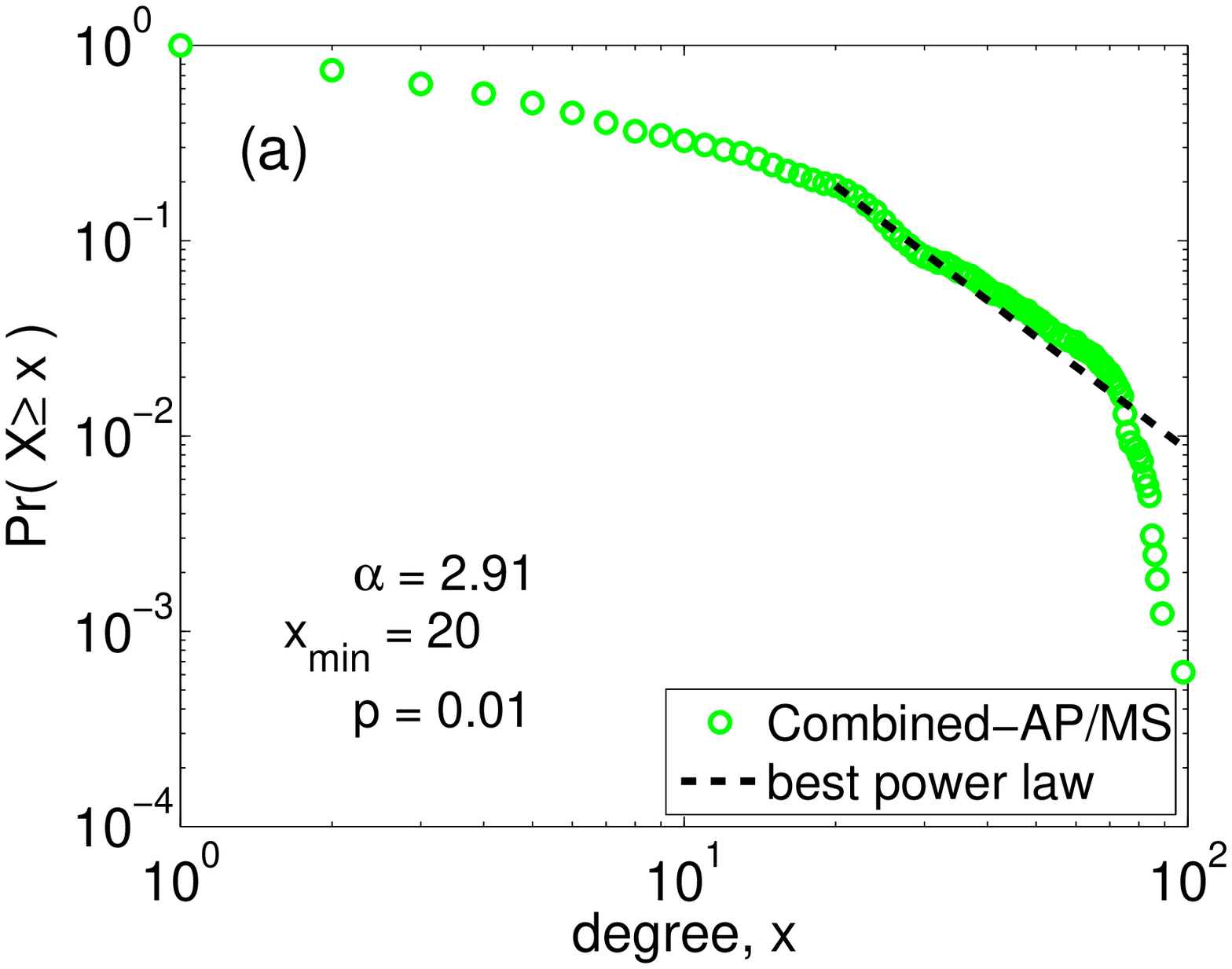} &
\includegraphics[scale=0.4]{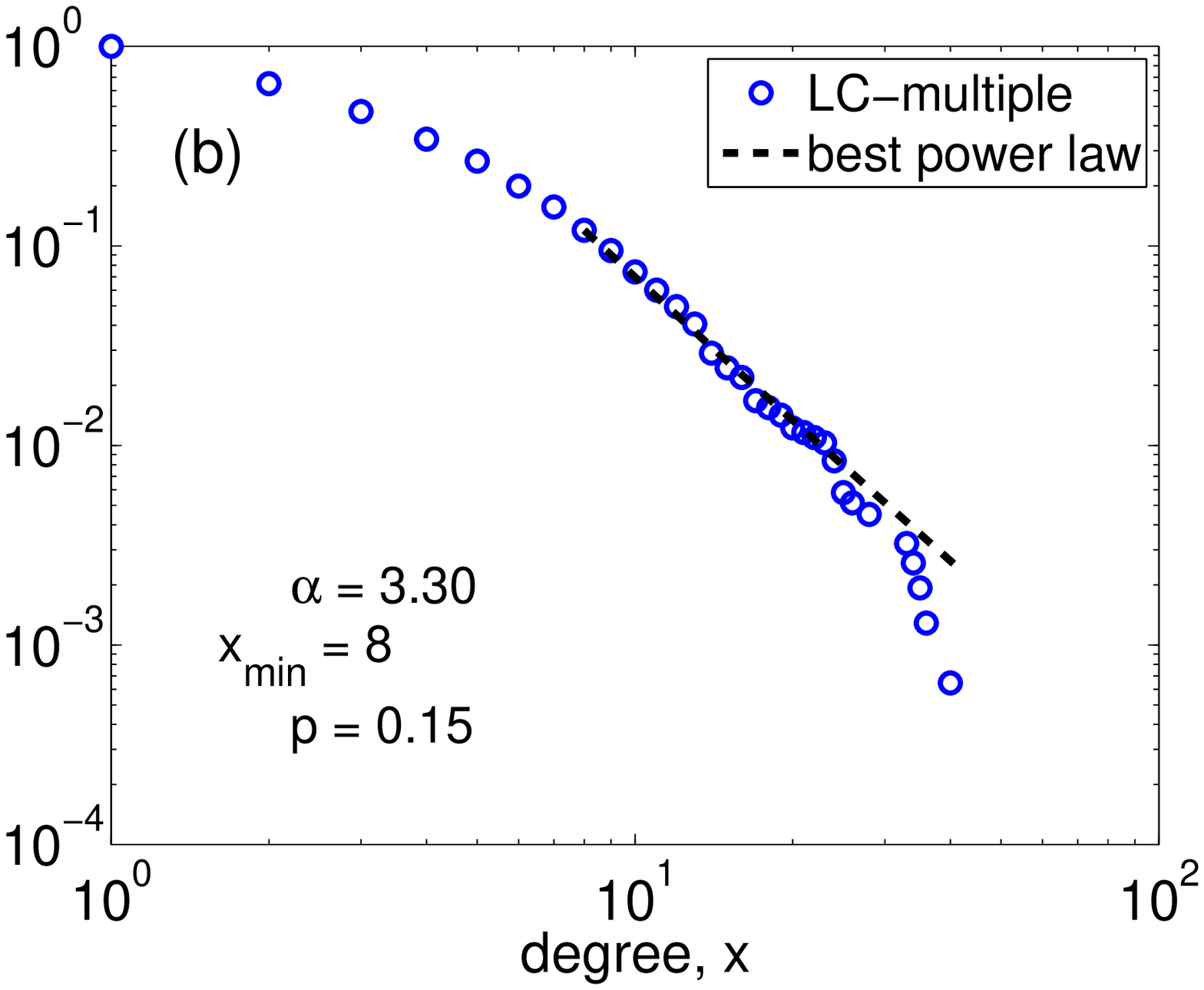} 
\end{tabular}
\end{center}
\caption{The Combined-AP/MS and LC-multiple data sets, shown as complementary cumulative distribution functions \mbox{$\Pr( X\geq x)$} on log-log axes, along with fits of the power-law hypothesis using the same methods as in Fig.~\ref{fig:S6A}c.}
\label{fig:S6B}
\end{figure*}

\begin{table*}
\begin{tabular}{l|cccc|cccc|c}
data set  & \multicolumn{4}{c}{regression} & \multicolumn{4}{c}{maximum likelihood} & ~support for~\\
& $\alpha$ & $x_{\rm min}$ & $n_{\rm tail}$ & $r^2$ & $\alpha$ & $x_{\rm min}$ & $n_{\rm tail}$ & $p$ ($\pm 0.03$)~ & power law\\
\hline
Y2H-union & ~2.4~ & 1 & ~2032~ & ~0.96~ & ~$2.9\pm 0.2$~ & ~$6\pm 1$~ & ~209~ & 0.95 & okay\\
Combined-AP/MS~ & 1.4 & 1 & 1621 & 0.96 & $2.9\pm 0.1$ & $20\pm 1$ & 309 & 0.01 & none\\
LC-multiple & 2.1 & 1 & 1552 & 0.92 & $3.3\pm 0.2$ & $8\pm 2$ & 186 & 0.15 & marginal
\end{tabular}
\caption{Power-law models for the three versions of the yeast interactome. In the first few columns, we quote the models given by Yu \etalp., which were derived using regression methods. In the second set, we give the best fits derived by maximum likelihood when the range of fit is allowed to vary (as in~\cite{clauset:etal:2007}), along with the corresponding $p$-value. In the final column, we state the support for the conjecture that the corresponding data follows a power-law distribution. In every case, the maximum likelihood power law is distinct and incompatible with that given by Yu \etalp., and only for the upper 10.3\% of the Y2H-union data does the power-law hypothesis have strong statistical support.
}
\label{table:1}
\end{table*}


The poor fit here is partly due to the way the power law was estimated from the data: linear regression for fitting the power law is known to produce biased results~\cite{goldstein:etal:2004,clauset:etal:2007} mainly by dramatically overweighting the large but rare events. Further, the canonical $r^{2}$ value, used to judge the quality of the regression, can easily be high even when the data are not power-law distributed~\cite{clauset:etal:2007}. In short, regression applied in this way makes assumptions about the model that are incompatible with the hypothesis being tested, and thus fails in uncontrolled ways.

A more reliable method for fitting a power-law distribution to data is the method of maximum likelihood~\cite{aldrich:1997,wasserman:2004}. Figure~\ref{fig:S6A}b shows a power-law model fitted to the entire range of data, while Figure~\ref{fig:S6A}c shows a power law fitted only to the upper range. The second of these requires choosing a point $x_{\min}$ where the power-law behavior starts, but this can be done using appropriate tools~\cite{clauset:etal:2007}. Applying the significance test to both models, we find that even the most likely power law is still a terrible explanation of the entire range of data ($p=0.00\pm0.01$), but that it's an entirely plausible explanation ($p=0.95\pm0.03$) of the 209 (10.3\%) most connected nodes, i.e., those with degree $x\geq6$. A corollary, however, is that the distribution of the degrees $1\leq x < 6$ is not a power law. (One possibility is that the low-degree data is power-law distributed in a way different from that of the high-degree nodes. Fortunately, the same tools we have already used can be readily adapted to test this hypothesis.)



Figure~\ref{fig:S6B} repeats the analysis of Figure~\ref{fig:S6A}c with the data for the alternative interactome maps. For these, we find that the Combined-AP/MS network's degree distribution cannot be considered to follow a power law ($p=0.01\pm0.03$), while the support for the LC-multiple network is marginally significant ($p=0.15\pm0.03$). In both cases, the best power-law models cover only the upper range of data, and power laws that cover the entire range have zero support ($p=0.00\pm0.03$). Table~\ref{table:1} summarizes the results of our reanalysis of the three networks.

As a brief aside, Yu \etal also conducted a model-comparison exercise to test whether a power-law distribution with or without an exponential cutoff was a better explanation of the data. For the same reasons given above, the regressions and $r^{2}$ values Yu \etal use cannot reliably determine which of these models is better. Fortunately, reliable tools for answering such a question do exist, e.g., a likelihood ratio test~\cite{vuong:1989,clauset:etal:2007}, although we do not apply them here.

With these results in hand, we can now make several novel conclusions about the structure of protein interactions in yeast, and generally clarify the results of Yu \etal First, the question of whether the yeast interactome's degree distribution is well-characterized by a power-law distribution is not yet settled. Yu \etal argue that the Y2H-union network is the most accurate map to date (more accurate than the Combined-AP/MS or LC-multiple versions), but here the statistics only support the notion that a power law is a plausible model of the degrees of a small fraction of the entire network (10.3\%, or the 209 most connected nodes). The LC-multiple network was presented as being a smaller, but generally high-quality, data set and here there is only marginal statistical support for a power-law degree distribution in the upper range. Notably, the two power laws for these networks are largely incompatible, with $\alpha_{\rm Y2H}=2.9\pm0.2$ versus $\alpha_{\rm LC}=3.3\pm0.2$. If the Y2H-union network were merely a more complete version of the LC-multiple network, a greater degree of overlap in these estimates would be expected. The implication is that there are significant differences between these networks that cannot be explained away by simple sampling arguments.

Further, the fact that the best power-law model of the Y2H-union data only explains the distribution of the upper 10.3\% of the node degrees implies that there is considerable structure in this network that remains to be explained. This structure may have evolutionary or functional significance, especially considering that its behavior is qualitatively different from that of the large-degree nodes and that it accounts for almost 90\% of the network. Additionally, the existence of the cross-over point at $x_{\min}=6$ from non-power-law to power-law behavior deserves a scientific explanation. In the additional structural analyses performed by Yu \etalp., the $x_{\min}$ value could serve as a principled threshold by which to quantitatively define ``high degree'' (see for instance~\cite{clauset:etal:2007b}). 

In closing, we note that there are many aspects of the Yu \etal study that seem entirely reasonable, and much of the paper concerns the experimental work done to construct the Y2H-union version of the yeast interactome. From this perspective, the paper pushes the field forward in a meaningful way, and the problems discussed here are a small part of a very large project.

On the other hand, the goal of constructing and analyzing the yeast interactome is to ultimately understand the mechanisms that create the observed patterns of interactions, and their implications for higher cellular functions. Scientific progress on these questions certainly depends on high quality experimental work, but it also depends on high quality statistical work: to get the theories right, we must also get the statistics right. Otherwise, we cannot know for sure what the data do and do not say.  For testing whether some data do or do not follow a power-law distribution, reliably accurate tools now exist, and their application can shed considerable light on the relevant scientific questions.

\vspace{-2mm}
\begin{acknowledgments}
\noindent 
Thanks go to C.\ R.\ Shalizi for encouraging me to post these thoughts on the arXiv.
\end{acknowledgments}


\begin{thebibliography}{99}
\bibitem{footnote:1}
Arguably, this level of understanding is still intermediate. A more complete understanding would incorporate knowledge about how genes interact and regulate each other, potentially through intermediate cellular processes.
\end{thebibliography}

\end{document}